\documentclass[final,1p]{elsarticle}
\usepackage{amssymb,amsmath,multirow,dcolumn,bm,float,graphicx,epstopdf}
\usepackage{calrsfs}  
\usepackage{epsfig}
\usepackage{soul}
\makeatletter

\usepackage{amssymb}
\usepackage{graphicx,color}
\usepackage[usenames,dvipsnames]{xcolor}
\usepackage{subfigure}

\usepackage{soul}

\usepackage{mathtools}
\usepackage{mathrsfs}
\usepackage{bm}
\usepackage{relsize}
\usepackage{indentfirst}

\usepackage{hyperref}
\hypersetup{
    unicode=false,          
    pdftoolbar=true,        
    pdfmenubar=true,        
    pdffitwindow=false,     
    pdfstartview={FitH},    
    pdfnewwindow=true,      
    colorlinks=true,       
    linkcolor=red,          
    citecolor=red,        
    filecolor=cyan,         
    urlcolor=magenta        
}

\newcommand{\nc}{\newcommand}
\nc{\lamnr}{\lambda_{nr}}
\nc{\lamr}{\lambda_{r}}
\nc{\lamk}{\lambda_{k}}
\nc{\gp}{g({\bp})}
\nc{\gpp}{g({\bp'})}
\nc{\gpz}{g({\bp''})}
\nc{\gps}{g^{*}({\bp})}
\nc{\gpzs}{g^{*}({\bp''})}
\nc{\gpps}{g^{*}({\bp'})}
\nc{\bxi}{{\bf \xi}}
\nc{\bp}{{\bf p}}
\nc{\bpp}{{\bf p'}}
\nc{\bpz}{{\bf p''}}
\nc{\bk}{{\bf k}}
\nc{\bkp}{{\bf k'}}
\nc{\bkz}{{\bf k''}}
\nc{\bPi}{{\bf \Pi}}
\nc{\bera}{\langle}
\nc{\ket}{\rangle}
\nc{\bq}{{\bf q}}
\nc{\bqp}{{\bf q'}}
\nc{\tpi}{\tilde{\pi}}
\nc{\bpi}{\boldsymbol \pi}
\nc{\btpi}{\tilde{\boldsymbol \pi}}
\nc{\andre}[1]{\textcolor{red}{#1}}

\journal{Physics Letters B}

\begin{document}
\begin{frontmatter}
\title{Three-boson stability for boosted interactions towards the zero-range limit}

\author[ITA]{K. Mohseni}
\ead{kmohseni@ita.br}
\author[ITA]{A. J. Chaves}
\author[Ceara]{D. R. da Costa}
\author[ITA]{T. Frederico}
\author[CSU,OU]{M. R. Hadizadeh}

\address[ITA]{Instituto Tecnol\'ogico de Aeron\'autica,  DCTA, 12228-900 S\~ao Jos\'e dos Campos,~Brazil}
\address[Ceara]{Departamento de F\'isica, Universidade Federal do Cear\'a, Caixa Postal 6030, Campus do Pici, 60455-900 Fortaleza, Cear\'a, Brazil}
\address[CSU]{College of Engineering, Science, Technology and Agriculture, Central State University, Wilberforce,
OH 45384, USA}
\address[OU]{Department of Physics and Astronomy, 
Ohio University, Athens, OH 45701, USA}

%
\begin{abstract}  
We study the three-boson bound-state mass and wave functions for ground and excited states within the three-body relativistic framework with Kamada and Gl\"ocke boosted potentials in the limit of a zero-range interaction.
 We adopt a nonrelativistic short-range separable potential, with Yamaguchi and Gaussian form factors, and drive them towards the zero-range limit by letting the form factors' momentum scales go to large values while keeping the two-body binding fixed. We show that the three-boson relativistic masses and wave functions are model-independent towards the zero-range limit, and the Thomas collapse is avoided, while the nonrelativistic limit kept the Efimov effect. Furthermore, the stability in the zero-range limit is a result of the reduction of boosted potential with the increase of the virtual pair center of mass momentum within the three-boson system. Finally, we compare the present results with Light-Front and Euclidean calculations.

\end{abstract}
\end{frontmatter}

The interpretation of novel results for the spectrum of hadron and multi-hadron states from Lattice Quantum Chromodynamics (LQCD) simulations serves as a new motivation to address relativistic few-body systems. 

In this context, it is known that two-body (2B) phase shifts can be extracted from the volume dependence of the spectrum by using the L\"uscher formula with two-particle quantization condition~\cite{Luscher:1986pf,Luscher:1990ux}. Beyond 2B, the spectra of multi-pion states in maximum isospin levels were computed by the NPLQCD collaboration~\cite{Beane_2008,Detmold:2008fn} in a finite volume more than a decade ago, and more recently, LQCD calculations were performed for the two- and three-pion finite-volume spectra for isospin $I=2$ and $3$, respectively~\cite{Horz:2019rrn}. 
The $I=3$ spectrum was calculated within a unitary three-dimensional framework for the two- and three-pion scattering amplitudes in a finite-volume discretized with periodic boundary conditions~\cite{Blanton:2019vdk} (see also~\cite{Mai:2018djl}).
Furthermore, the finite-volume energy spectrum of the $K^-K^-K^-$ system was also obtained from LQCD calculations~\cite{Alexandru:2020xqf}. 

A field theory on a four-dimensional (4D) lattice is defined with periodic (anti-periodic) boundary conditions in the case of bosonic (fermionic) fields, which includes the time direction as well (see e.g.~\cite{Gattringer:2010zz}). In correspondence, the 4D finite-volume formulation of the Faddeev-like equations in Euclidean space was recently implemented~\cite{Frederico:2021wej} in the
4D three-boson Faddeev-Bethe-Salpeter (FBS) equation for the contact interaction~\cite{Frederico:1992np}. 

An important issue that permeates these continuum treatments of the relativistic three-boson system is its stability in the limit of a zero-range interaction, as it is known that the Thomas collapse~\cite{Thomas:1935zz,Coutinho:1995zv} is present in the nonrelativistic three-boson system with contact potentials. The stability
was shown by solving the Light-Front (LF) reduction of the FBS equation truncated at the valence state~\cite{Carbonell:2002qs, Karmanov:2003qk}.  The stability of the three-boson system was also shown by solving the 4D FBS equation for the contact interaction in Euclidean space~\cite{YDREFORS2017131}. It was then confirmed the relative importance of the implicit three-body interactions of relativistic origin~\cite{Karmanov:2009bhn}, which is missing in the formulation of the valence state integral equation. 

This context motivates us to study the stability of the three-boson system in the limit of contact interactions within other frameworks to formulate the relativistic Faddeev equations, for example, using boosted potentials~\cite{KAMADA2007119}.
The boost concept comes from the moving 2B subsystem in the rest frame of the three-particle system. One approach for calculating 2B boosted $T-$matrices is solving the relativistic Lippmann-Schwinger equation for boosted 2B potentials. At the 2B level, the relativistic 2B potentials are designed to preserve the 2B observables for bound and scattering states. Kamada and Gl\"ockle have shown that relativistic and boosted potentials can be obtained directly from nonrelativistic potentials by solving a quadratic equation using an iterative scheme~\cite{KAMADA2007119}. Once the boosted potential is obtained, the 2B $T-$matrix in the rest frame of the three-body system can be computed as required for the kernel of the relativistic Faddeev equations~\cite{PhysRevC.90.054002,hadizadeh2020three}.

The above formulation will be explored  to study the three-boson stability separable interactions (Yamaguchi-type and Gaussian-type) driven to the zero-range limit. Our goal is to solve the relativistic Faddeev equations, and obtain both the binding energies and associated wave functions, and study their properties when the interaction range is driven towards zero.

Before presenting our study in detail, we should remark that the relativistic Faddeev approach with boosted interactions is not based on field theory, but it uses a relativistic version of the phenomenological 2B potential. It belongs to one of the three forms of relativistic dynamics proposed by Dirac in 1949~\cite{Dirac1949}: the instant form, the LF form, and the point form. The different forms of dynamics are characterized by the number of kinematical and dynamical boosts, in correspondence with the generators of the Poincar\'e group. The most commonly used forms of dynamics are the instant and LF forms, and we restrict ourselves to those ones. The kinematical boosts keep the initial state hyper-surface invariant and do not contain the interaction, while the dynamical ones depend on the interaction. In the instant form, six out of ten generators of the Poincar\'e group are kinematical, while four are dynamical and contain the interaction. In the LF form, seven generators are kinematical, and three are dynamical. This form corresponds to the one with a maximal number of kinematical generators (for a thorough discussion of this form of dynamics applied to nuclear few-body systems, see e.g. Ref.~\cite{CarbonellPREP1998}). 
The relativistic framework developed by Kamada and Gl\"ockle is within the class of the instant form dynamics. Their development applied to few-nucleon systems was designed to keep the relativistic deuteron binding energy and nucleon-nucleon phase shifts unaltered from the results obtained with the nonrelativistic calculations. Furthermore, it provides energy states with good angular momentum quantum numbers, as usual in the nonrelativistic frameworks. In this way, the relativistic Faddeev approach with boosted potential can be viewed as one possible and practical implementation of the instant form of dynamics. We will also quantitatively illustrate the difference between the Kamada and Gl\"ockle instant form framework outcomes and the results from LF and field theoretical models in the limit of zero-range interactions.

{\it Relativistic Faddeev approach with boosted interactions.} The relativistic Faddeev equations for the bound state of three identical particles were recently derived in momentum space as a function of relativistic Jacobi momentum vectors \cite{PhysRevC.90.054002,hadizadeh2020three}. A partial wave projection of relativistic Faddeev equation in an $s-$wave channel is given by
\begin{equation}
\psi(p,k) = 4\pi\, G^r_0(p,k)\,  
 \int_0^\infty 
  \hbox{d}  k'  {k'_2}^2 
  \int_{-1}^1\hbox{d} x'\, 
N(k,k',x') \ T_k(p,\tpi;\epsilon) \ \psi(\pi, k'),
  \label{Faddeev_PW}
\end{equation}
where $p$ and $k$ are the relativistic Jacobi momenta, $G^r_0(p,k)=[M_t-\omega_k(p)-\Omega(k)]^{-1}$ is the relativistic free propagator, where $M_t = E_t + 3m$ is the 3B mass eigenvalue, $m$ is the mass of each particle, $\omega_k(p)=\sqrt{\omega^2(p)+k^2}$, $\omega(p)=2\sqrt{p^2+m^2}$, and $\Omega(k) = \sqrt{m^2+k^2}$.
The definitions of the shifted momentum $\pi$ and $\tpi$, and the remaining quantities are given in~\ref{app}.
In the nonrelativistic limit where the momenta are much smaller than the masses, the Jacobian function $N$, given in~\ref{app}, reduces to one. In addition, the relativistic Jacobi momenta $p$ and $k$ reduce to the corresponding nonrelativistic Jacobi momenta, and similarly, the shifted momentum arguments $\tpi$ and $\pi$ reduce to the corresponding nonrelativistic ones of Ref.~\cite{elster1999three}.
The boosted 2B transition matrix $T_{k}(p,p';\epsilon)$ for 2B subsystem energies $\epsilon=M_t- \Omega(k)$ is obtained from relativistic Lippmann-Schwinger equation as
\begin{equation} \label{eq.t_boost_PW}
T_{k}(p,p';\epsilon)=V_{k}(p,p')+ 4\pi \int dp'' p''^2  \, \frac{V_{k}(p,p'')}
{ \epsilon - \omega_k(p'') }  
\,T_{k}(p'',p';\epsilon).
\end{equation}
The matrix elements of boosted potential $V_{k}$ can be obtained directly from the nonrelativistic potential $V_{nr}$, by solving a quadratic integral equation \cite{KAMADA2007119}
\begin{equation}
V_k(p,p') + \frac{4\pi}{\omega_k(p)+\omega_k(p')}
\int d p'' p''^2 \, V_k(p,p'') \, V_k (p'',p')  = \frac{4m \, V_{nr}(p,p')
}{\omega_k(p)+\omega_k(p')}.
\label{eq.Vk_Vnr}
\end{equation}
An important physical property of the boosted potential, clear in Eq.~\eqref{eq.Vk_Vnr}, is the damping with the increase of the momentum of the spectator particle $k$. This behavior corresponds in practice to an effective three-body repulsive effect, working in the ultraviolet region (UV), which balances the attraction at the short range. We will illustrate these properties when studying the three-boson bound state when driving the potential range to zero.

{\it Results.} In the following, we present our numerical results for the solution of relativistic Faddeev integral equation~\eqref{Faddeev_PW} for boosted potentials obtained from one-term separable nonrelativistic potentials, which are generally defined as $V_{nr}(p,p') = \lambda_{nr} \, g(p) \, g(p') $, where $\lambda_{nr}$ is the potential strength and $g(p)$ is the form factor in the momentum basis. In this work we use two models of separable potentials, Yamaguchi-type form factor $g(p)=1/(p^2+\beta^2)$ \cite{PhysRev.95.1628} and the Gaussian form factor $g(p)=\exp{(-p^2/\Lambda^2)}$ \cite{deltuva2011universality}.
\begin{figure}[hbt]
\centering
\includegraphics[width=0.72\linewidth]{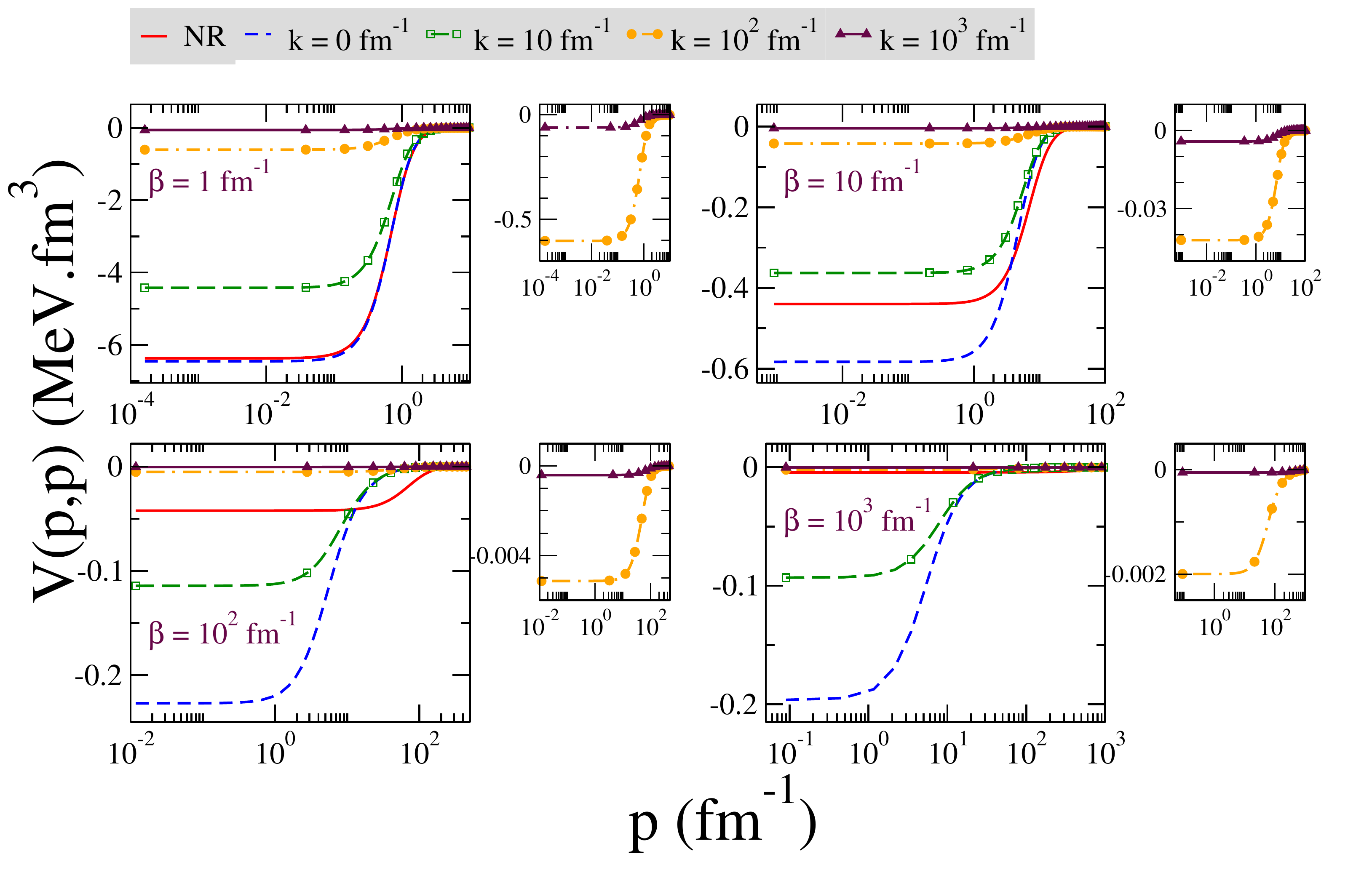}
\includegraphics[width=0.72\linewidth]{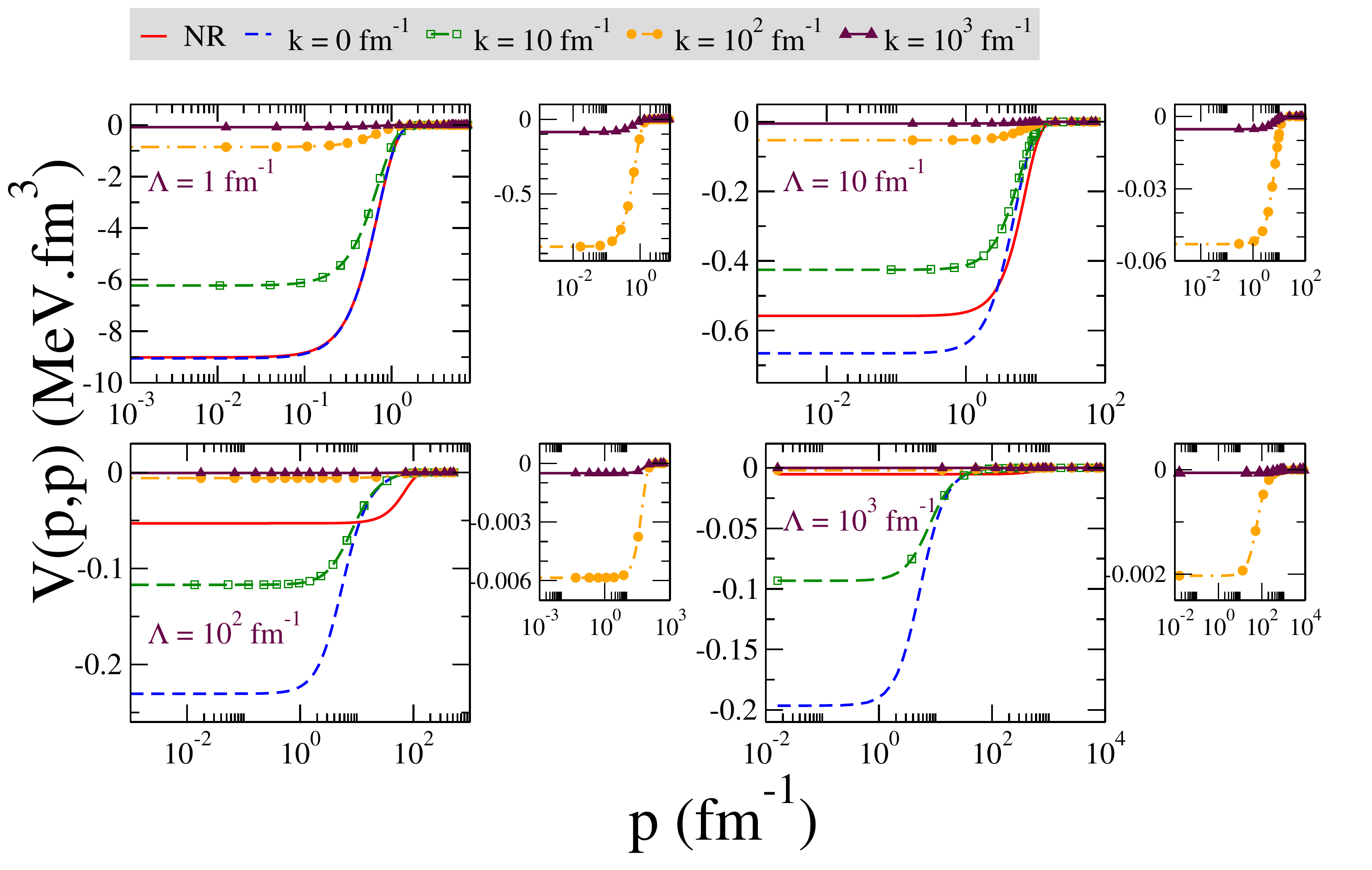}
\caption{The diagonal matrix elements of nonrelativistic and boosted potentials calculated for a wide range of form factor parameters $\beta$ for Yamaguchi-type (upper panel) and $\Lambda$ for the Gaussian potentials (lower panel) obtained with different boost momentum $k$.}
\label{NR_boost}
\end{figure}

The inputs for the solution of relativistic Lippmann–Schwinger equation \eqref{eq.t_boost_PW} are the matrix elements of boosted potentials $V_k(p, p')$ which can be obtained directly from nonrelativistic interaction $V_{nr}(p, p')$ by solving the integral Eq. \eqref{eq.Vk_Vnr} using an iterative scheme proposed by Kamada and Gl\"ockle \cite{KAMADA2007119} and successfully implemented in a three-dimensional scheme \cite{hadizadeh2017calculation,hadizadeh2021nnrelativistic}. The iteration starts with the initial guess
\begin{equation}\label{eq.V0}
V_{k}^{(0)}(p,p')=\frac{4m \, V_{nr}(p,p') }{\omega_k(p)+\omega_k(p')}\, ,
\end{equation}
and continues to reach convergence in the matrix elements of the boosted potential with a relative error of $10^{-16}$~MeV~fm$^3$ at each set points $(p,p')$. 
In Fig.~\ref{NR_boost}, we show the diagonal matrix elements of boosted potentials with different values of boost momentum $k$ calculated for a wide range of form factor parameters $\beta$ for Yamaguchi-type and $\Lambda$ for the Gaussian potentials.
As one can see, the boosted potentials are getting smaller by increasing the boost momentum $k$.

\begin{figure}[thb]
\centering
\includegraphics[width=.450\linewidth]{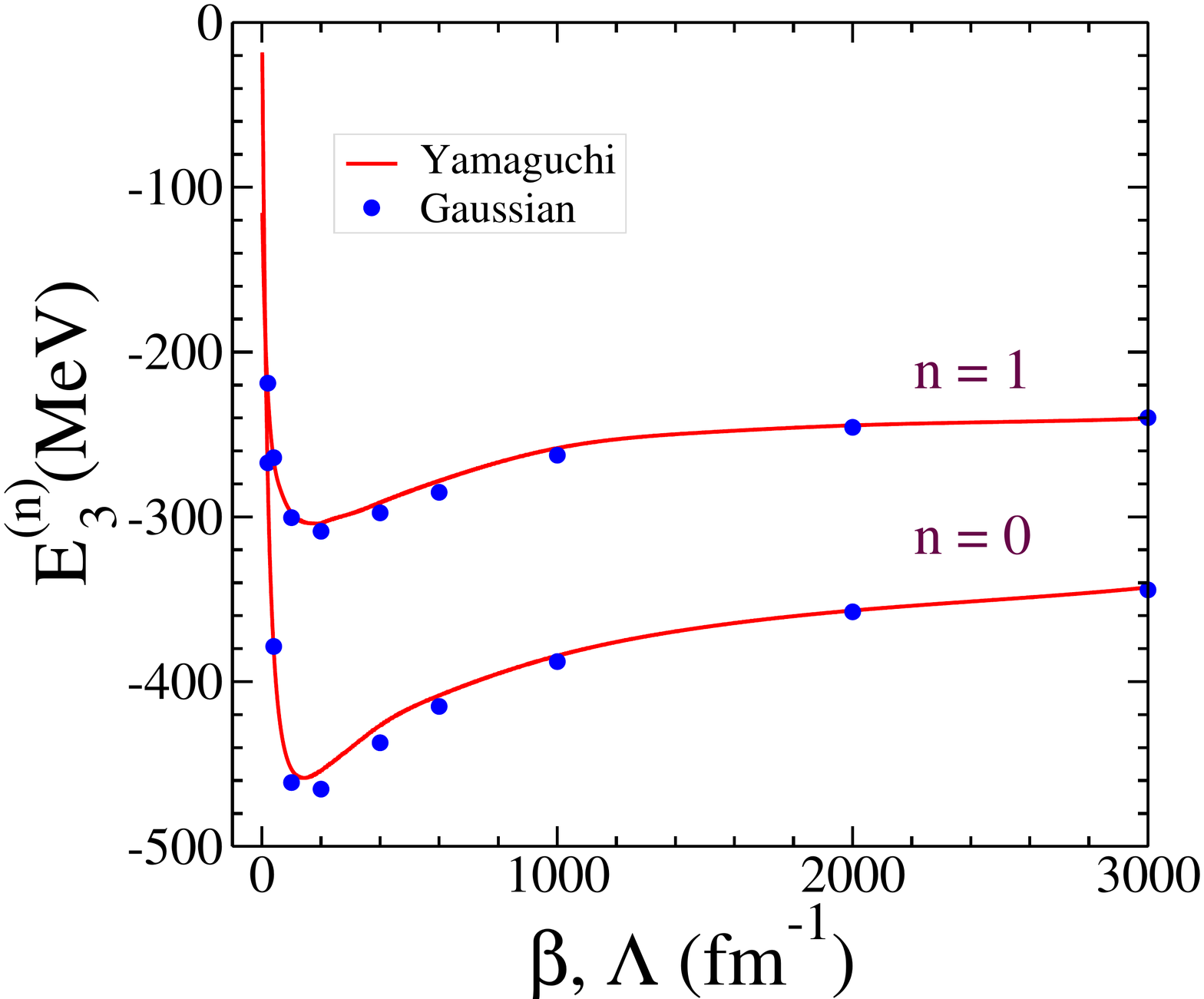}
\includegraphics[width=.450\linewidth]{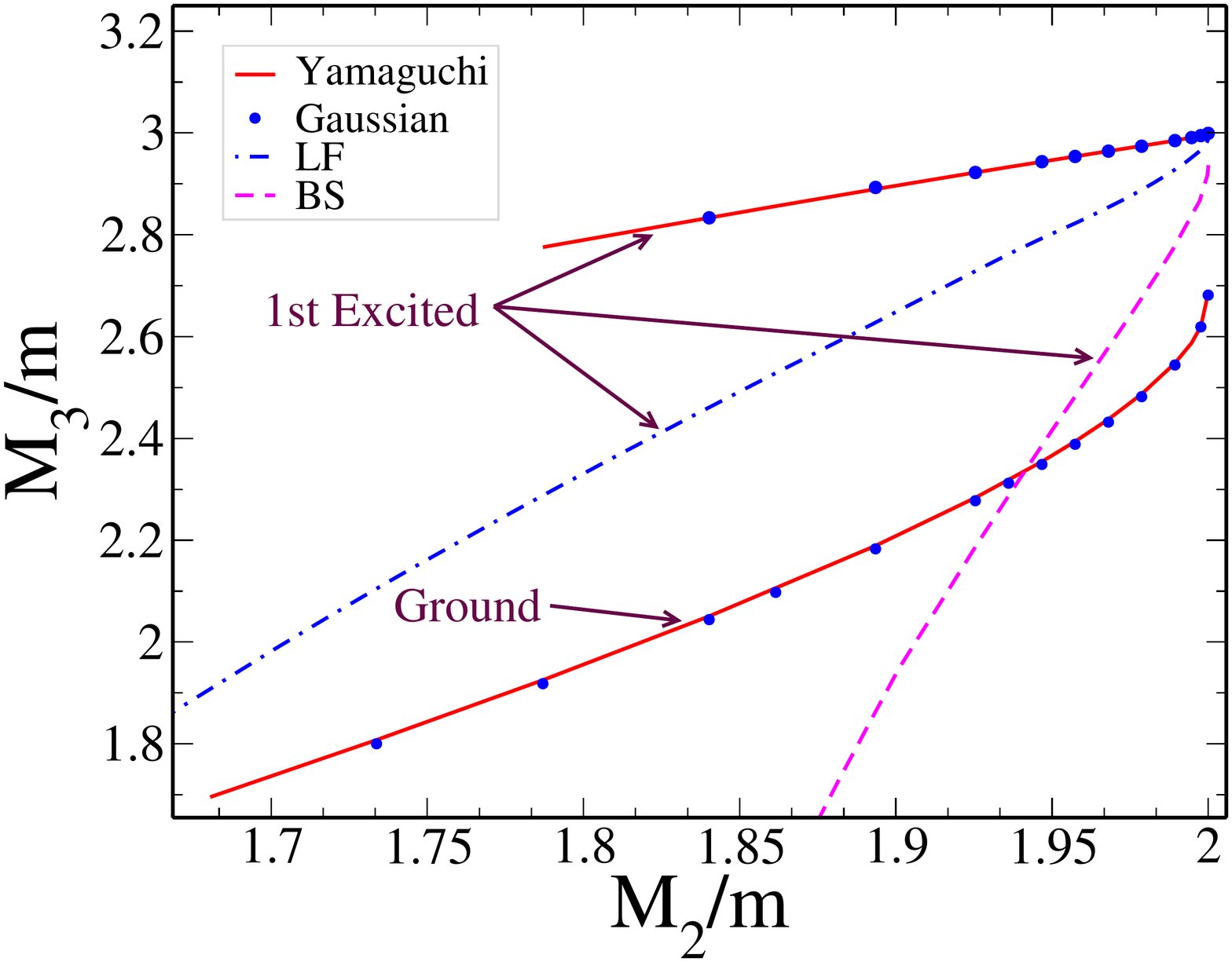}
\caption{Left panel: the ground and first excited state binding energies as a function of the form factor parameters $\beta$ (solid line for Yamaguchi-type potential) and $\Lambda$ (full circles for the Gaussian potential) obtained for a fixed 2B binding energy of $-2.225$ MeV. The first excited state binding energies are multiplied by a factor of $50$. Right panel: the value of $M_3/m$ as a function of $M_2/m$ calculated from the three-body ground and first excited states obtained with form factor parameters $\beta=\Lambda=2000$~fm$^{-1}$ for Yamaguchi-type (solid lines) and Gaussian potentials (full circles). For comparison, our results with the Bethe-Salpeter (dashed line) and Light-Front (dash-dotted line) zero-range calculations of Ref. \cite{YDREFORS2017131} are added to the plot.}
\label{fig_E3_Lambda_beta}
\end{figure}

By having the matrix elements of boosted potentials $V_{k}(p,p')$, we solve the relativistic Lippmann-Schwinger integral equation~(\ref{eq.t_boost_PW}) to calculate fully off-shell boosted $T-$matrices $T_{k}(p,p';\epsilon)$ for 2B subsystem energies $\epsilon=M_t- \Omega(k)$ dictated by the boost momentum $k$. Then, by solving the integral equation (\ref{Faddeev_PW}) with the Lanczos technique (see Appendix C2 of Ref. \cite{PhysRevA.85.023610}), we obtain relativistic 3B binding energies $E_{t}$ and Faddeev components $\psi(p,k)$ for ground and excited states. We use the Gauss-Legendre quadratures with hyperbolic plus linear mapping for Jacobi momenta and linear mapping for angle variables to discretize continuous momentum and angle variables~\cite{hadizadeh2017calculation}. The cutoffs of Jacobi momenta and the distribution of their mesh points strongly depend on the potential form factor parameters $\beta$ and $\Lambda$. 

In the left panel of Fig.~\ref{fig_E3_Lambda_beta}, we show three-boson ground and first excited state binding energies as a function of the potential range parameters $\beta$ and $\Lambda$, for the Yamaguchi and Gaussian form factors, respectively, with a nonrelativistic 2B binding energy kept at -2.225~MeV, namely, the deuteron binding energy. For $\beta$ and $\Lambda$ increasing, the three-boson system follows the Thomas collapse, with the binding energy $\propto -\beta^2$ and $\propto -\Lambda^2$, happening in our examples up to values around 200~fm$^{-1}$ or corresponding to a momentum of 0.4~GeV/c comparable to the nucleon mass. Then, the boost effects take place and stabilize the system through an induced repulsion that tends to counterbalance the singular behavior of the collapse with the binding energy reaching a plateau, regardless of the short-range potential model, suggesting that a well-defined zero-range limit exists for the Gl\"ockle-Kamada boosted potentials within the relativistic three-body framework.

In the right panel of Fig.~\ref{fig_E3_Lambda_beta}, we present three-body and single-particle mass ratio $M_3/m$ as a function of the 2B and single-particle mass ratio $M_2/m$, obtained for three-body ground and first excited states using large form factor parameters $\beta = \Lambda = 2000$ fm$^{-1}$. The plot presents the mass in units of the particle mass, which can be compared with previous calculations using the Bethe-Salpeter (BS) and Light-Front equations~\cite{YDREFORS2017131}.
As one can see in the right panel of Fig.~\ref{fig_E3_Lambda_beta}, our results with large form factor parameters for both Yamaguchi-type and Gaussian potentials reveal a universal behavior. 
Our numerical results for three-boson first excited bound state mass obtained from the boosted potentials with large form factor parameters, i.e. $\beta = 2000$~fm$^{-1}$ and $\Lambda = 2000$~fm$^{-1}$, show a weaker attraction when compared to the results obtained with the LF and BS frameworks~\cite{YDREFORS2017131}.
First, we observe that, in the 2B bound state region, both the LF and BS approaches have an unphysical three-body ground state with $M_3^2<0$, which are possible as  the homogeneous integral equations, only depend on $M_3^2$~\cite{YDREFORS2017131}. The physical ``ground'' state with $0<M^2_3<(m+M_2)^2$ from the solution of the LF and BS equations are indeed an excited state, and in this way, they are denoted in the right panel of the figure. 
\begin{figure}[thb]
\centering 
\includegraphics[width=.49\linewidth]{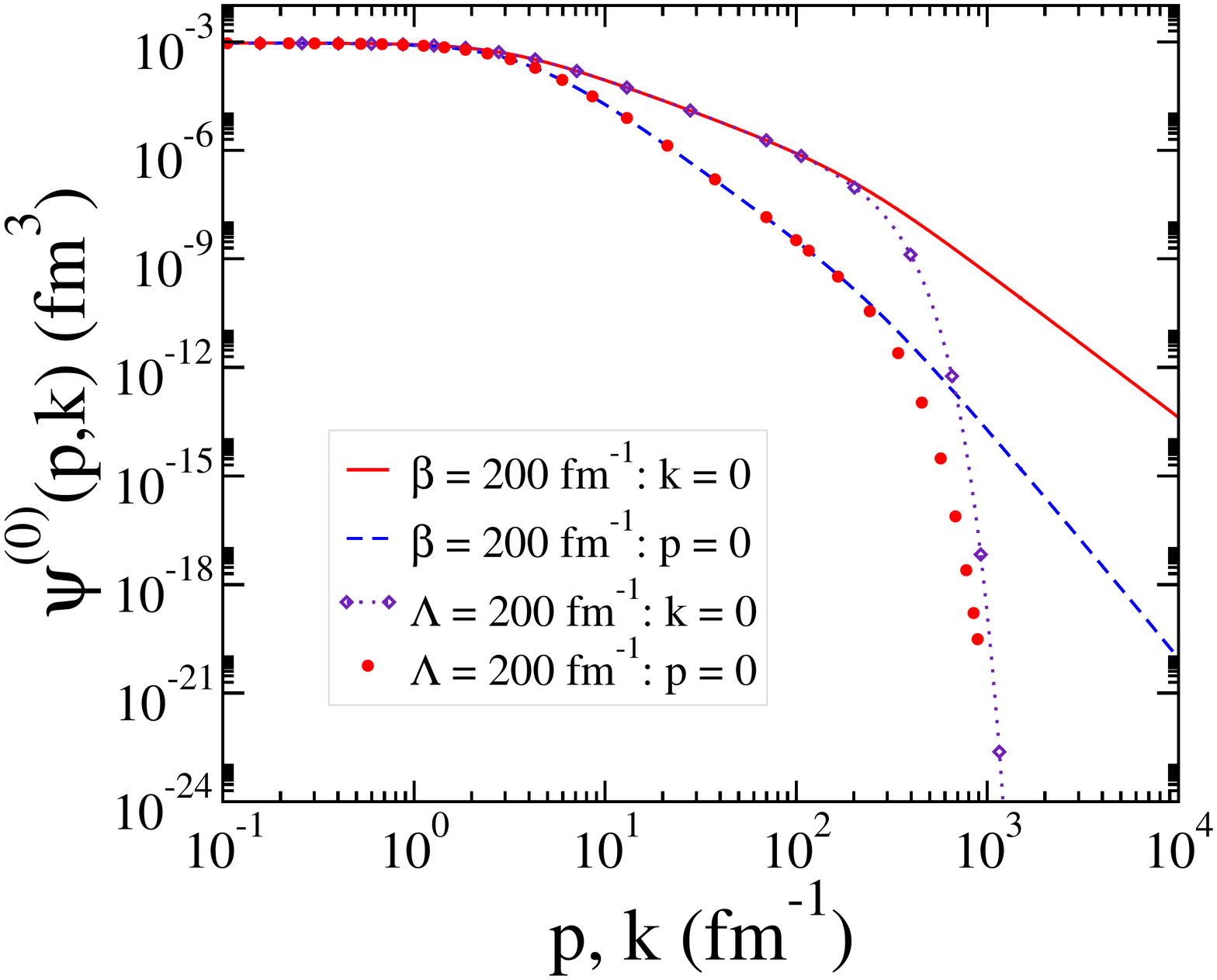}
\includegraphics[width=.49\linewidth]{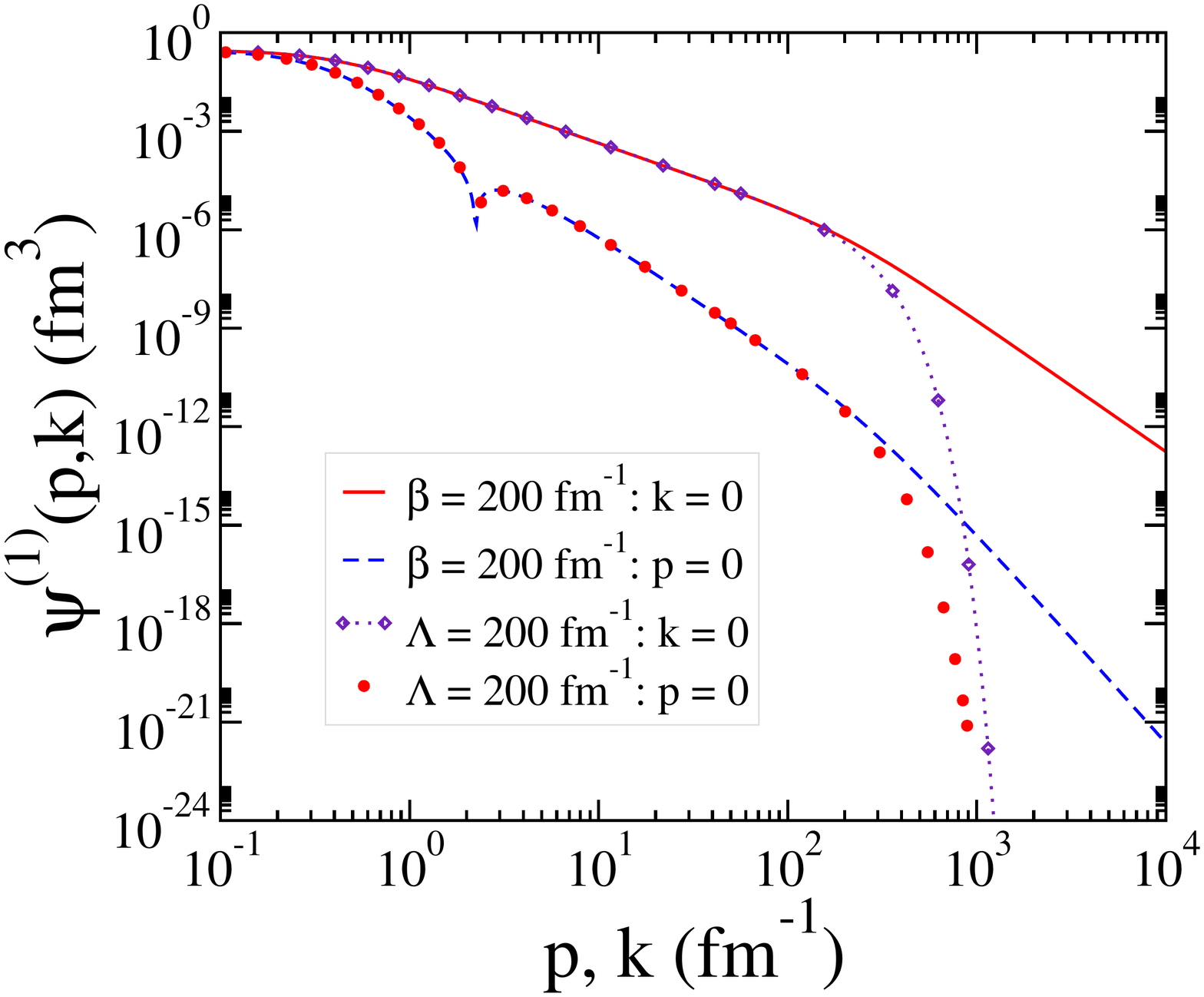} \\
\includegraphics[width=.49\linewidth]{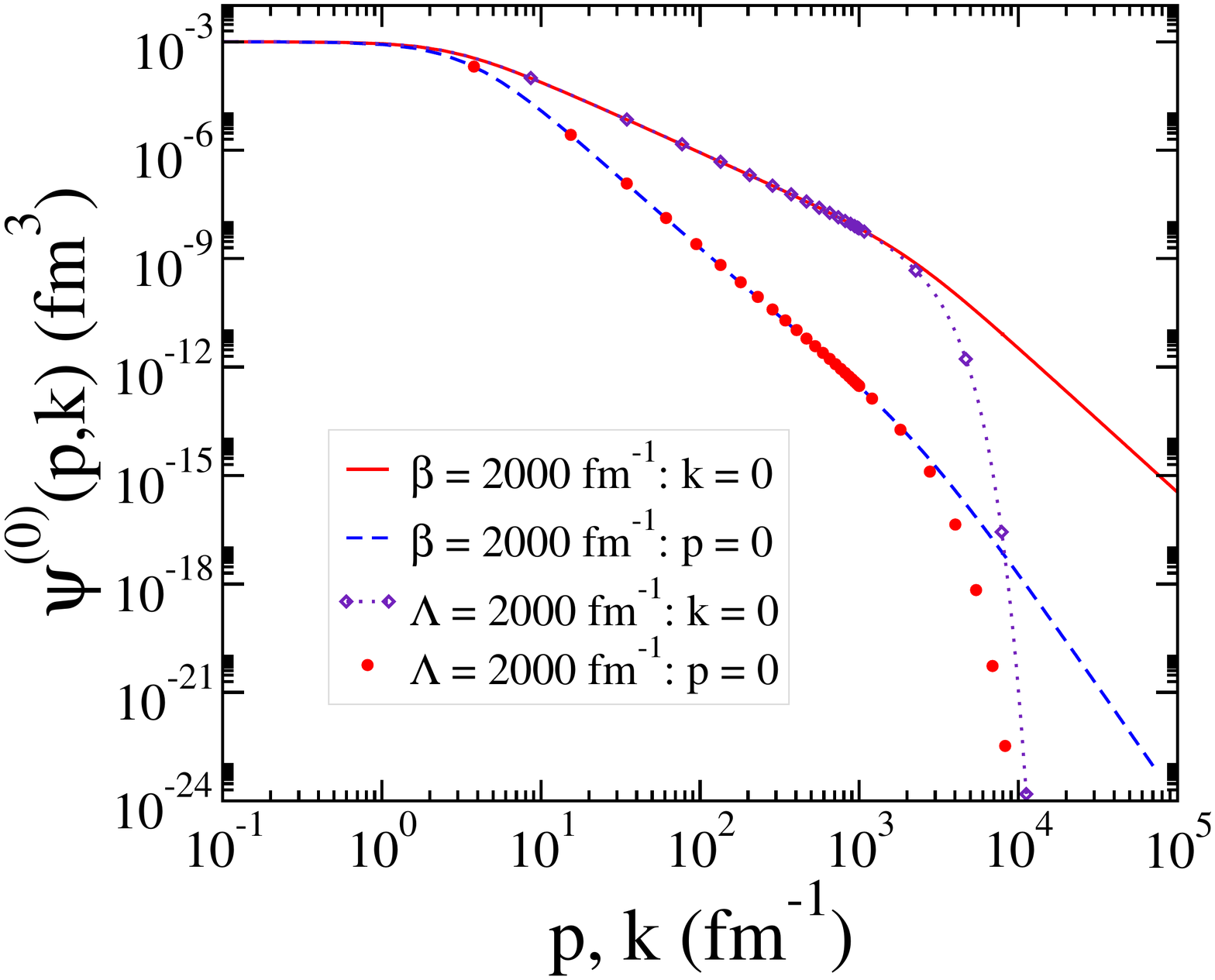} 
\includegraphics[width=.49\linewidth]{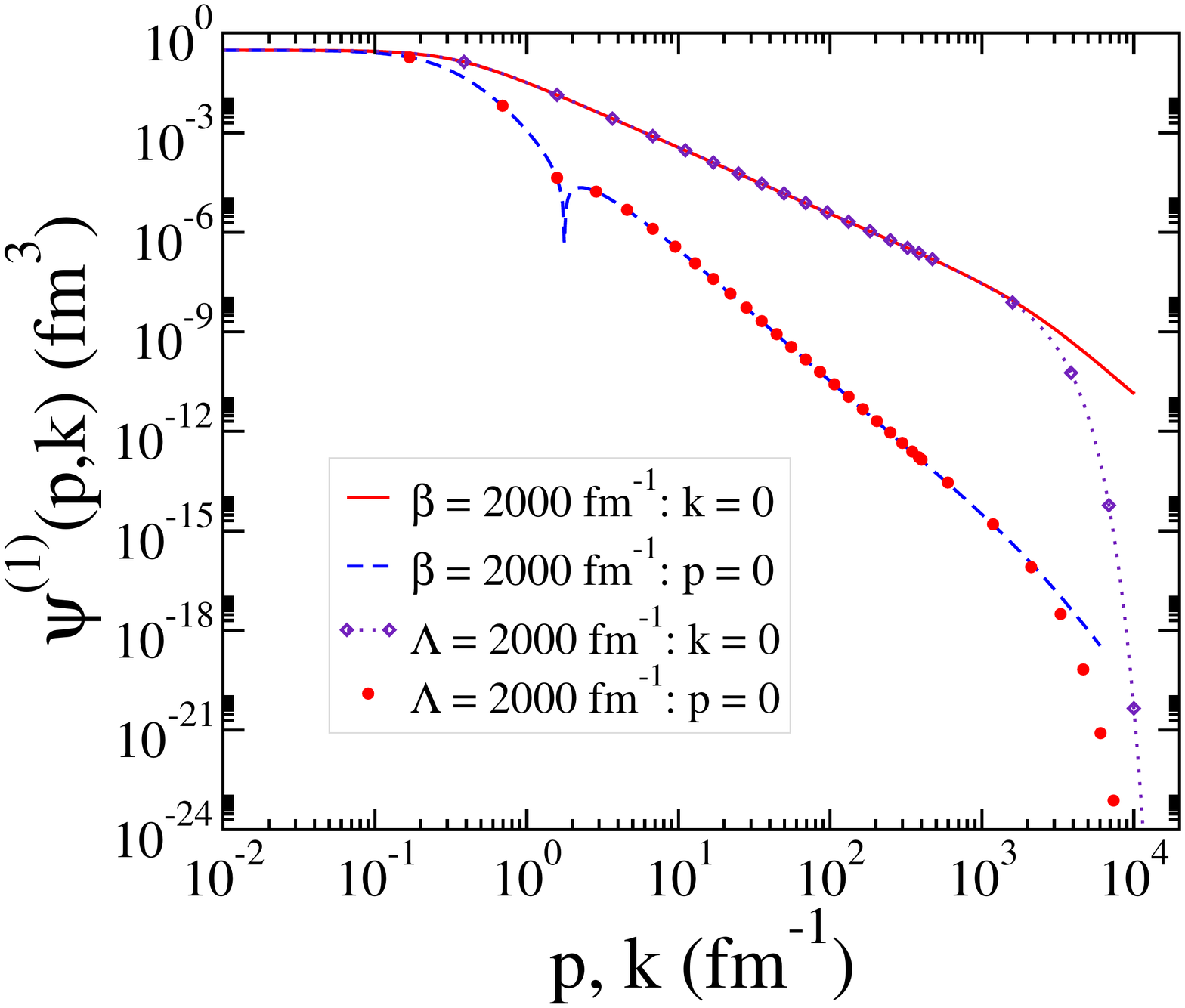} 
\par
\caption{Relativistic Faddeev components for three-body ground $\psi^{(0)}(p,k)$ and first excited $\psi^{(1)}(p,k)$ states calculated for Yamaguchi-type potential with form factor parameters (lines) $\beta=200,~2000$ fm$^{-1}$ and Gaussian potential with form factor parameters (symbols) $\Lambda=200,~2000$ fm$^{-1}$.}
\label{Faddeev_components}
\end{figure}

The difference between the LF and BS approaches is the attractive three-body effective interactions appearing in the BS approach due to the implicit inclusion of an infinite set of LF Fock-components, which is missing in the LF equation where the truncation is made at the valence level~\cite{YDREFORS2017131}.
What is noticeable is that the boosted potential brings less attraction to the three-body system, and furthermore in the UV region is much less attractive due to the softening of the relativistic potential owing the boost coming with the solution of Eq.~\eqref{eq.Vk_Vnr}. This effect can be appreciated by noticing the slowest decrease of $M_3$ by decreasing $M_2$ for the boosted potential calculation with respect to BS and LF results.
The softening in the decrease of the three-body mass with the increase of the 2B binding in the adopted relativistic framework, as seen in the right panel of Fig.~\ref{fig_E3_Lambda_beta}, suggests that the increase of the binding turns the three-body bound state more compact, which forces the system to explore the UV region, where the boosted potential becomes weak.
On the other hand, the 2B amplitude considered in the LF and BS equations is much less damped in the UV region, presenting a $\sim 1/\log (k)$ behavior (cf. Eq.~(2) in Ref.~\cite{YDREFORS2017131}),
quite soft compared to the decrease of the boosted potential matrix elements, with $\sim 1/k$, as the non-linear term is not relevant for $k>>m$ attaining very large values, as the driving term of the iterative solution for the boosted potential, Eq.~\eqref{eq.V0}, follows $V^{(0)}_k(p,p')|_{k>>m}\propto 1/k$. Thus, this discussion is indeed indicating a weaker kernel of the Faddeev equation provided by the boosted potential with respect to the LF and BS ones.

Finally, in Fig.~\ref{Faddeev_components}, we show the Faddeev component of the wave functions corresponding to the ground state, $\psi^{(0)}(p,k)$, and first excited one, $\psi^{(1)}(p,k)$, as a function of Jacobi momenta $p$ and $k$, for two sets of small and large potential form factor parameters $\beta$ and $\Lambda$. The corresponding binding energies are listed in Table~\ref{Lamda_beta_energy}.
The first striking observation is the universality, i.e., the model independence, of  $\psi^{(0)}(p,k)$ and $\psi^{(1)}(p,k)$ up to momenta about $\beta$ or $\Lambda$. For larger momenta, both potential models clearly show different decay behaviors, and the Gaussian model shows, as expected, the faster damping of the Faddeev component, while the Yamaguchi one still seems a power-law behavior.

The interesting aspect seen in Fig.~\ref{Faddeev_components} is the model independence of the power-law damping of the wave function happening both for the third particle spectator momentum $m<k<\beta (\Lambda)$, and the two-particle relative one $m<p<\beta (\Lambda)$. Furthermore, the power-law behavior in the region of momenta between $\sim m$ and $\sim \beta(\Lambda)$ is independent of the excitation state of the three-boson system for both momenta. The fit of the power-law functions in the region $m<k<\beta (\Lambda)$, scales with $\psi^{(0)}(p,k=0)\sim p^{-1.9}$ and $\psi^{(0)}(p=0,k)\sim k^{-3.7}$.

\begin{table}[thb]
\caption{Relativistic three-body ground and first excited state binding energies for Faddeev components shown in Fig. \ref{Faddeev_components}. The 2B binding energy is $-2.225$ MeV.}
\begin{center}
\begin{tabular}{ccccccccccccccc}
\hline
Potential parameter (fm$^{-1}$) & $E_t^{(0)}$ (MeV) & $E_t^{(1)}$ (MeV) \\
\hline
\multicolumn{3}{c}{Yamaguchi-type potential} \\
\hline
$\beta=200$  & $-454.1$ & $-6.076$ \\
$\beta=2000$ & $-353.7$ & $-4.891$ \\
\hline
\multicolumn{3}{c}{Gaussian potential} \\
\hline
$\Lambda=200$ & $-465.2$ & $-6.176$ \\
$\Lambda=2000$ & $-357.7$ & $ -4.914$ \\
\hline
\end{tabular}
\end{center}
\label{Lamda_beta_energy}
\end{table}%
We should observe that the power-law property of the wave function component in the relative momentum $p$ is somewhat expected, as the relativistic propagator behavior dominates it at large $p$ values. 
On the other hand, the power-law in $k$ is in contrast with the log-periodic behavior characteristic of the Thomas collapsed states in the limit of the zero-range interaction for the nonrelativistic three-boson  system~\cite{skornyakov1956jetp,Danilov1961}. On the other hand, the Efimov effect~\cite{Efimov:1970zz}, when the 2B binding is let to zero, is kept by the Faddeev equations with boosted potentials.

{\it Summary.} In this work, we solved the
relativistic three-boson bound state problem
within the Kamada and Gl\"ocke framework of building the boosted potential in the limit of a zero-range interaction. The starting point is a nonrelativistic short-range separable potential, with Yamaguchi and Gaussian form factors, which are driven to the contact interaction by letting their momentum scales to large values in comparison with the boson mass scale, while the 2B binding energy was kept fixed.

The solutions of the relativistic three-boson Faddeev equation with the boosted potentials are stable, and the three-boson relativistic masses and wave functions are finite and model-independent towards the limit of the zero-interaction. 
The Thomas collapse is avoided, while the Efimov physics for large scattering lengths are kept. The boosted potential provides an effective repulsive short-range effect that decreases the interaction intensity and guarantees the stability of the relativistic three-boson system towards the zero-range limit. 

We found that the Faddeev equation with the boosted potential provides a much weaker attraction to bind the three-boson system than to what was found by solving the LF and BS ones~\cite{YDREFORS2017131}. That effective weaker attraction can be partially traced back to the more strong damping of the boosted potential, and equally well the 2B T-matrix with the spectator momentum with a characteristic behavior $\sim 1/k$, while within the LF and BS formulations, the 2B amplitude is damped as $\sim 1/\log k$,
quite soft compared to the decrease of the boosted potential matrix elements. Furthermore, we have confirmed that the stability of the three-boson system with boosted potentials in the zero-range limit is accompanied by a universal power-law behavior of the wave function component when the momenta are large. This is in contrast with the log-periodic behavior characteristic of the Thomas collapsed states in the limit of the zero-range interaction.

We want to add that the adopted framework may find applications beyond hadron physics due to the renewed interest in few-body complexes in condensed matter, motivated by the recently synthesized two dimensional materials (see, e.g., \cite{berkelbach2013theory}). For example, in the case of monolayers of gapped honeycomb materials like Transition Metal Dichalcogenides~\cite{wang2018colloquium}, anisotropic 2D semiconductors~\cite{trions2D} and hexagon Boron Nitride~\cite{zhang2017two}, the dispersion relation is hyperbolic~\cite{kormanyos2015k,ferreira2019excitons}, thus allowing the use of relativistic frameworks in 2D, but requiring the consideration of the spinor property of the quasi-particles~\cite{chaves2017excitonic}. These possible applications are left for future studies. 

{\it Acknowledgments.} T.F. thanks E. Ydrefors for the  discussions.
This work is a part of the project INCT-FNA proc. No. 464898/2014-5.
K.M. acknowledges a PhD scholarship from the Brazilian agency CNPq (Conselho Nacional de Desenvolvimento Cient\'ifico e Tecnol\'ogico).
T.F. acknowledges the CNPq Grant No. 308486/2015-3  and Funda\c{c}\~ao de Amparo \`a Pesquisa do Estado de S\~ao Paulo (FAPESP) under Thematic Project 2017/05660-0. 
The work of M.R.H. was supported by the National Science Foundation under Grant No. NSF-PHY-2000029 with Central State University. This work was supported by CNPq Grant No. 400789/2019-0.

\appendix

\section{Definitions}\label{app}
The definitions and notation shown in this appendix follow Refs.~\cite{PhysRevC.90.054002,hadizadeh2020three}. The shifted momentum arguments appearing in Eq.~\eqref{Faddeev_PW} are given by
\begin{eqnarray} 
\tpi &=& 
\sqrt{\frac{1}{4}C^2(k,k',x')k^2+k'^2+C(k,k',x')kk'x'},\\
 \cr
\pi &=&
\sqrt{k^2+\frac{1}{4}C^2(k',k,x')k'^2+C(k',k,x')kk'x'},
\label{eq.shifted_momenta}
\end{eqnarray}
where the permutation coefficients $C(k,k',x')$ are defined as
\begin{equation} 
C(k,k',x') = 1+ \frac{ \Omega(k')- 
\Omega(\mathcal{K}) }
{ \Omega(k') + \Omega(\mathcal{K} ) + 
\sqrt{\bigl ( \Omega(k') 
+ \Omega (\mathcal{K}) \bigr)^2-k^2} }.
\label{eq.C}
\end{equation}
As a note, permutation coefficients $C$ reduces to one in the nonrelativistic limit.
The Jacobian function $N(k,k',x')$ is defined as
\begin{eqnarray}
N(k,k',x') &=&
\left( 
\frac{4\Omega(k') \Omega(\mathcal{K})}
{ \sqrt{ \biggl(\Omega(k') +\Omega(\mathcal{K}) \biggr)^2-k^2 }
\biggl( \Omega(k') +\Omega(\mathcal{K}) \biggr)
} \
\right ) ^{-\frac{1}{2}}
\cr
&\times&
\left( 
\frac{4\Omega(k) \Omega(\mathcal{K})}
{ \sqrt{ \biggl (\Omega(k) +\Omega(\mathcal{K}) \biggr)^2-k'^2 } \
\biggl( \Omega(k) +\Omega(\mathcal{K}) \biggr)
}
\right ) ^{-\frac{1}{2}},
\label{eq.N}
\end{eqnarray}

where $\mathcal{K}=\sqrt{k^2+k'^2+2kk'x'}$.


\end{document}